\documentclass{PoS}

\usepackage[utf8]{inputenc}
\usepackage[normalem]{ulem}
\usepackage[english]{babel}
\usepackage{tabularx}
\usepackage{array}
\usepackage{graphics}
\usepackage[pdftex]{graphicx}
\usepackage{psfrag}
\usepackage{epsfig}
\usepackage{amsmath}
\usepackage{amssymb}
\usepackage{bbold}
\usepackage{setspace}
\usepackage{rotating}
\usepackage{colortbl}
\usepackage{longtable}
\usepackage{slashed}
\usepackage{braket}
\usepackage{lineno}
\makeatletter
\usepackage{textcomp}
\usepackage[usenames,dvipsnames]{xcolor}
\usepackage{relsize}
\usepackage{cite}

\usepackage{ae}
\usepackage{mathtools}
\usepackage{amsfonts}
\usepackage{amssymb}
\usepackage{bbold}

\usepackage{enumerate}

\usepackage{bbm}
\usepackage{bm}
\usepackage{enumerate}
\usepackage{amsmath}

\usepackage{amssymb}
\usepackage{pifont}
\newcommand{\xmark}{\ding{55}}%

\newcommand{\bea}{\begin{eqnarray}}
\newcommand{\eea}{\end{eqnarray}}
\newcommand{\beq}{\begin{equation}}
\newcommand{\eeq}{\end{equation}}
\newcommand{\ec}{\end{center}}
\newcommand{\bc}{\begin{center}}

\newcommand{\gev}{{\rm GeV}}

\newcommand{\pdir}{p\kern -5.2pt\raise 0.2ex\hbox {/}}

\newcommand{\vdir}{v\kern -5.75pt\raise 0.15ex\hbox {/}}
\newcommand{\kdir}{k\kern -5.75pt\raise 0.15ex\hbox {/}}
\newcommand{\epsdir}{\epsilon\kern -5.0pt\raise 0.15ex\hbox {/}}
\newcommand{\bvdir}{\bar{v}\kern -5.75pt\raise 0.15ex\hbox {/}}
\newcommand{\Ddir}{D\kern -7.75pt\raise 0.20ex\hbox {/}}
\newcommand{\Adir}{A\kern -7.75pt\raise 0.20ex\hbox {/}}
\newcommand{\ldir}{l\kern -5.0pt\raise 0.2ex\hbox{/}}
\newcommand{\varepsdir}{\varepsilon\kern -5.5pt\raise 0.15ex\hbox{/}}

\title{Lepton flavor (universality) violation in $B$-meson decays}

\ShortTitle{Lepton flavor (universality) violation in $B$-meson decays}

\author{\speaker{Olcyr Sumensari}\\
        Dipartimento di Fisica e Astronomia `G. Galilei', Universit\`a di Padova, Italy\\
        Istituto Nazionale Fisica Nucleare, Sezione di Padova, I-35131 Padova, Italy\\
        E-mail: \email{olcyr.sumensari@pd.infn.it}}

\abstract{Even though the LHC searches did not unveil the new physics particles so far, LHCb hints towards deviations from lepton flavor universality in exclusive decays based on the transition $b\to s\ell\ell$. In this proceeding, we present a new leptoquark model that can explain both $R_{K}^{\mathrm{exp}}<R_{K}^{\mathrm{SM}}$ and $R_{K^\ast}^{\mathrm{exp}}<R_{K^\ast}^{\mathrm{SM}}$ via loop effects, consistent with observations made by LHCb. We discuss the main predictions of this scenario that can be tested experimentally, which include the bounds on lepton flavor violating decays  $\mathcal{B}(Z\to\mu\tau) \lesssim \mathcal{O}(10^{-7})$ and $\mathcal{B}(B\to K\mu\tau)\lesssim \mathcal{O}(10^{-9})$.}

\FullConference{The European Physical Society Conference on High Energy Physics\\
		5-12 July, 2017\\
		Venice}

\begin{document}

\section{Introduction}

One of the most intriguing results obtained so far at the Large Hadron Collider (LHC) is the indication of the lepton flavor universality violation (LFUV) in semileptonic $B$ meson decays. First, from the measured partial branching fractions of $B\to K\ell^+\ell^-$, in the window of $q^2 \in [1,6] \ \gev^2$, the LHCb Collaboration in Ref.~\cite{Aaij:2014ora} reported
\begin{align}
\label{exp:RK}
R_K=  \dfrac{\mathcal{B}(B^+\to K^+\mu\mu)}{\mathcal{B}(B^+\to K^+ ee)} \Bigg{\vert}_{q^2\in(1,6)~\mathrm{GeV}^2}=0.745\pm^{0.090}_{0.074}\pm 0.036\,,
\end{align}
\noindent which lies $2.6\sigma$ below the the clean SM prediction $R_K^{\mathrm{SM}}=1.00(1)$~\cite{Hiller:2003js}. This observation of LFUV was recently corroborated by the most recent LHCb results in two $q^2$-bins~\cite{Aaij:2017vbb},
\begin{align}
\label{exp:RKstar}
R_{K^\ast}^{\mathrm{low}} &= \frac{ \mathcal{B}( B \to K^\ast \mu \mu)_{q^2\in [0.045,1.1]~\mathrm{GeV}^2}}{\mathcal{B}( B \to K^\ast e e)_{q^2\in~[0.045,1.1]\mathrm{GeV}^2}} = 0.660 \pm^{0.110}_{0.070} \pm 0.024 \, ,\nonumber\\[0.6em]
R_{K^\ast}^{\mathrm{central}} &= \frac{ \mathcal{B}( B \to K^\ast \mu \mu)_{q^2\in [1.1,6]~\mathrm{GeV}^2}}{\mathcal{B}( B \to K^\ast e e)_{q^2\in [1.1,6]~\mathrm{GeV}^2}} = 0.685 \pm^{0.113}_{0.069} \pm 0.047 \,,
\end{align}
\noindent thus again $\approx (2.2-2.4) \sigma$ below the Standard Model (SM) predictions~\cite{Hiller:2003js}. When combined in the same fit, these results amount to a discrepancy with respect to the SM at the $4\sigma$ level~\cite{DAmico:2017mtc}. Since the hadronic uncertainties largely cancel out in $R_{K^{(\ast)}}$, if confirmed, these would be an unambiguous manifestation of NP. 

Several models have been proposed to accommodate $R_{K^{(\ast)}}^{\mathrm{exp}}<R_{K^{(\ast)}}^{\mathrm{SM}}$ through the Wilson coefficients $C_{9(10)}^{\mu\mu}$, see Ref.~\cite{Becirevic:2017jtw} and references therein. Among those, the models postulating the existence of low energy leptoquark (LQ) states are of particular interest as we will discuss in the following.  

\section{Tree-level leptoquark models for $b\to s\ell\ell$}

LQs are colored states mediating interactions between quarks and leptons, which can be a scalar or a vector field and which may come as a $SU(2)_L$-singlet, -doublet or -triplet~\cite{Buchmuller:1986zs,Dorsner:2016wpm}. Among these scenarios, the ones invoking vector LQs are not renormalizable and become problematic when computing the loop-induced processes, such as $\tau\to\mu\gamma$ and the $B_s\to \overline{B_s}$ mixing amplitude.~\footnote{In other words, the predictivity of these scenarios is compromised unless a renormalizable and gauge invariant ultraviolet completion is explicitly specified. See Ref.~\cite{DiLuzio:2017vat} for a first proposal of UV completion with light vector LQs.} For this reason, we will focus on scatar LQ scenarios and assume that the SM is extended by only one LQ state. 

In Table~\ref{tab:lq-classification-RK}, we classify by their SM representation the scalar LQ states that can modify $R_{K^{(\ast)}}$ through tree-level contributions to $b\to s \mu\mu$~\cite{Becirevic:2016oho,Hiller:2017bzc}.  From this table, we see that only the scenario with a scalar triplet $(\bar{\mathbf{3}},\mathbf{3})_{1/3}$ can accommodate both $R_K^{\mathrm{exp}}<R_K^{\mathrm{SM}}$ and $R_{K^\ast}^{\mathrm{exp}}<R_{K^\ast}^{\mathrm{SM}}$ through the effective coefficients $C_9^{\mu\mu}=-C_{10}^{\mu\mu}$. Nonetheless, this model violates baryon number via the dangerous diquark couplings, which can induce the proton decay at tree-level~\cite{Dorsner:2016wpm}. In this case, an additional symmetry is needed to forbid these couplings from destabilizing the proton (see Ref.~\cite{Dorsner:2017ufx} for an example in the framework of grand unification). Notice also that scenarios invoking the $(\mathbf{3},\mathbf{2})_{1/6}$ state, originally proposed to explain $R_K^{\mathrm{exp}}<R_K^{\mathrm{SM}}$, became disfavored after the recent observation of $R_{K^\ast}^{\mathrm{exp}}<R_{K^\ast}^{\mathrm{SM}}$ by LHCb, which disagrees with its predictions \cite{Becirevic:2015asa}.

\begin{table}[htb!]
\renewcommand{\arraystretch}{1.3}
\centering
\begin{tabular}{|c|ccccc|}
\hline 
\quad $(SU(3)_c,SU(2)_L)_{U(1)_Y}$  & BNC & Interaction & Eff.~Coefficients & $R_K/R_{K}^{\mathrm{SM}}$ & $R_{K^\ast}/R_{K^\ast}^{\mathrm{SM}}$	 \\ \hline\hline
$(\boldsymbol{\bar{3},3})_{1/3}$	&   \xmark & $\overline{Q^C} i \tau_2 \boldsymbol{\tau}\cdot \boldsymbol{\Delta} L$	&$C_9=-C_{10}$	& $<1$ & $<1$ \\  
$(\boldsymbol{\bar{3},1})_{4/3}$	&   \xmark	& 	$\overline{d^C_R} \boldsymbol{\Delta}\ell_R$ &$(C_9)^\prime=(C_{10})^\prime$	&	$\approx1$ & $\approx 1$\\ 
$(\boldsymbol{3,2})_{7/6}$	&  \checkmark	& $\overline{Q}\boldsymbol{\Delta}\ell_R$	&	$C_9=C_{10}$ 	&	$>1$ & $>1$ \\  
$(\boldsymbol{3,2})_{1/6}$	&  \checkmark	&	$\overline{d_R}\widetilde{\boldsymbol{\Delta}}^\dagger L$	& $(C_9)^\prime=-(C_{10})^\prime$	& $<1$ & $>1$ \\  
\hline
\end{tabular}
\caption{\label{tab:lq-classification-RK}\small \sl List of LQ states classified by their SM quantum numbers which can modify the transition $b\to s \mu\mu$ at tree-level. The conservation of baryon number (BNC), the interaction term and the corresponding Wilson coefficients are also listed along with the prediction for $R_K$. Couplings to electrons are set to zero.}
\end{table}

In the following we will argue that the model $(\mathbf{3},\mathbf{2})_{7/6}$ can also be used to explain both $R_K^{\rm exp} < R_K^{\rm SM}$ and $R_{K^\ast}^{\rm exp} <R_{K^\ast}^{\rm SM}$ provided the tree-level contributions to $b\to s \ell\ell$ ($\ell=e,\mu$) are absent~\cite{Becirevic:2017jtw}. In our implementation, the relevant Wilson coefficients are only induced at loop-level and satisfy $C_9^{\mu\mu}=-C_{10}^{\mu\mu}<0$, making both $R_K$ and $R_{K^\ast}$ smaller than in the SM.

\section{$(3,2)_{7/6}$ or $R_2$ leptoquark model}

The so-called $R_2$ model involves a doublet of scalar leptoquarks with hypercharge $Y=7/6$. The general Yukawa Lagrangian for this model reads
\begin{align}
\label{eq:slq1}
\begin{split}
\mathcal{L}_{\Delta^{(7/6)}} &= (g_R)_{ij} \,\overline{Q}_i{\boldsymbol\Delta}^{(7/6)}\ell_{Rj}+
                                                           (g_L)_{ij}
                                                           \, \overline{u}_{Ri} { \widetilde{\boldsymbol\Delta}}^{(7/6) \dagger} L_{j}+ \mathrm{h.c.}\,,
\end{split}
\end{align}
where $g_{L,R}$ are the matrices of Yukawa couplings, that we take to be
\begin{equation}
\label{eq:YC}
g_{L} = \left( \begin{matrix}
  0 & 0 & 0\\
  0 & g_{L}^{c \mu} & g_{L}^{c \tau}\\
  0 & g_{L}^{t \mu} & g_{L}^{t \tau}
\end{matrix}\right), \qquad g_{R} = \left( \begin{matrix}
  0 & 0 & 0\\
  0 & 0 &0\\
  0 & 0 & g_{R}^{b \tau}
\end{matrix}\right), \qquad V g_{R} = \left( \begin{matrix}
  0 & 0 &  V_{ub} g_{R}^{b \tau}\\
  0 & 0 & V_{cb} g_{R}^{b \tau}\\
  0 & 0 & V_{tb} g_{R}^{b \tau}
\end{matrix}\right), 
\end{equation}
which is the main peculiarity of our model. The conjugate $SU(2)_L$ doublet is defined by $ { \widetilde{\boldsymbol\Delta}}^{(7/6)} = i \sigma_2{\boldsymbol\Delta}^{(7/6)}$ and we use $Q_i = [(V^\dagger u_L)_i\; d_{Li}]^T$ and
$L_i = [(U\nu_L)_{i}\; \ell_{Li}]^T$ to denote the quark and lepton doublets, in which $V$ and $U$ are the
Cabibbo-Kobayashi-Maskawa (CKM) and the Pontecorvo-Maki-Nakagawa-Sakata (PMNS) matrices, respectively. 

The above choice of Yukawa couplings, and in particular $g_R^{s\mu}=g_R^{b\mu}=0$, means that the contributions of the leptoquark to the transition $b\to s\mu\mu$ can only appear at loop-level. The only diagrams contributing are box diagrams which give rise to the following Wilson coefficients~\cite{Becirevic:2017jtw}
\begin{align}\label{eq:C9new}
C_9^{\ell_1\ell_2}=-C_{10}^{\ell_1\ell_2}= \sum_{u,u^\prime \in \{u,c,t\}} {V_{ub} V_{u^\prime s}^\ast\over V_{tb} V_{t s}^\ast } g_L^{u^\prime\ell_1} \left( g_L^{u \ell_2}\right)^\ast \mathcal{F}(x_u, x_{u^\prime})\,,
\end{align}
where $x_{i}= m_{i}^2/m_W^2$, and the loop function reads, 
\begin{align}
\mathcal{F}(x_u, x_{u^\prime})= {\sqrt{x_u x_{u^\prime}} \over 32\pi\alpha_\mathrm{em}}  &\biggl[ {  x_{u^\prime} (  x_{u^\prime} - 4) \log  x_{u^\prime}\over ( x_{u^\prime}-1) (x_u- x_{u^\prime})( x_{u^\prime}-x_\Delta)}
+ { x_u (  x_u - 4) \log  x_u\over ( x_u-1) (x_{u^\prime} - x_u )( x_u -x_\Delta)} 
\biggr.\nonumber\\[2.ex]
&\biggl.  - { x_\Delta (x_\Delta -4) \log x_\Delta \over (x_\Delta -1) (x_\Delta - x_u)( x_\Delta - x_{u^\prime} ) } 
\biggr]\,.
\end{align}
\noindent This closes our discussion of the $R_2$ model with our particular setup specified by the structure of the $g_{L,R}$ matrices, as given in Eq.~\eqref{eq:YC}. We shall now discuss the phenomenological implications of this scenario.

\section{Flavor constraints and predictions}

To assess the viability of this model, we performed a scan of parameters by varying the couplings in Eq.~\eqref{eq:slq1} within the perturbativity limit, $|g_{L,R}^{i,j}|<\sqrt{4\pi}$, and the LQ mass $m_\Delta$ in ther interval $m_\Delta \in (0.65,4)~\mathrm{TeV}$. The allowed parameters are then confronted with several phenomenological constraints of which the most relevant ones are: (i) the branching ratios for $\mathcal{B}(B_s\to\mu\mu)$ and $\mathcal{B}(B\to K\mu\mu)_{\mathrm{high}~q^2}$, (ii) the branching ratios $\mathcal{B}(Z\to\ell\ell)$, with $\ell=\mu,\tau$, (iii) limits on $\mathcal{B}(B\to K\nu\nu)$, (iv) the experimentally established $(g-2)_\mu$, and (v) the bounds on $\mathcal{B}(\tau\to\mu\gamma)$, cf.~Ref.~\cite{Becirevic:2017jtw} for details. We also impose the available LHC limits for pair produced LQs decaying into $\Delta^{(2/3)}\to t\nu$ and $\Delta^{(5/3)}\to t\tau$ \cite{Aad:2015caa}, which are reinterpreted the pattern for Yukawa couplings we consider, cf.~Eq.~\eqref{eq:slq1}.

\begin{figure}[ht!]
\centering
\includegraphics[width=0.5\linewidth]{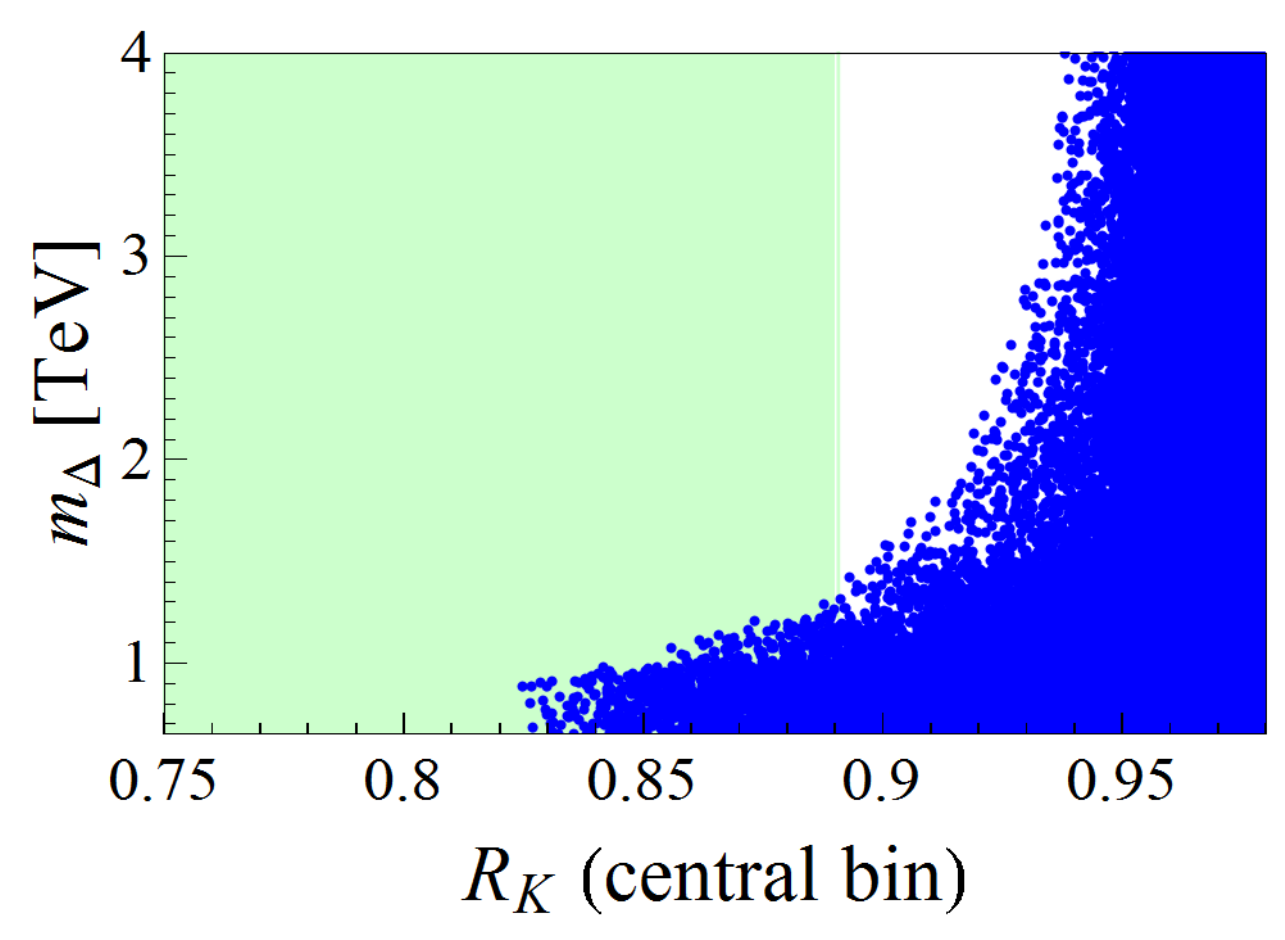}~\includegraphics[width=0.5\linewidth]{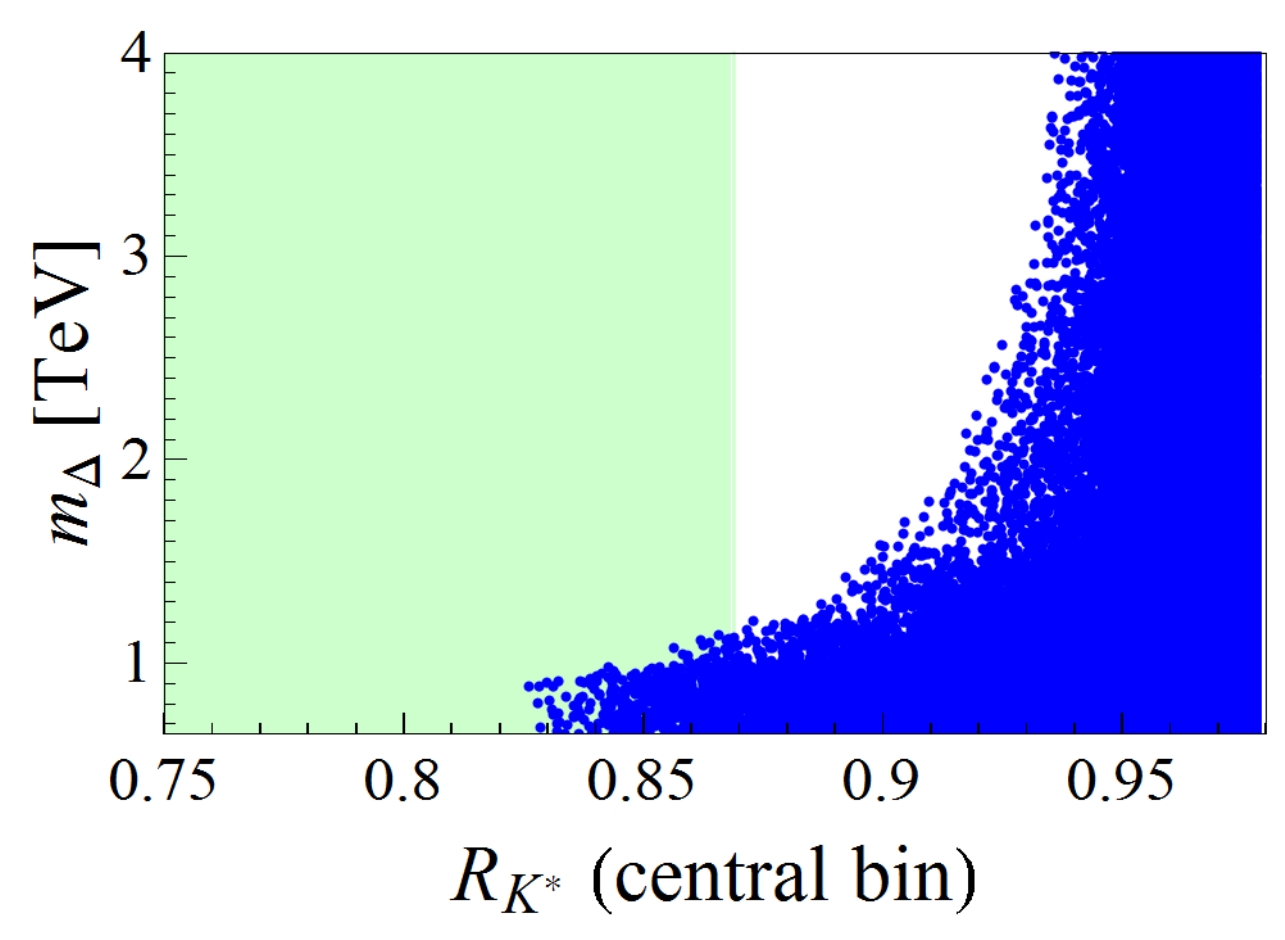}\\
\caption{\small \sl Results of our scan of parameters consistent with all constraints discussed in the previous section in which the leptoquark mass $m_\Delta$ is varied too. We see that the $1.5 \sigma$  consistency requirement with the values of LHCb for $R_K$ and $R_{K^\ast}$ in the central $q^2$-bin (shaded area) results in $m_\Delta < 1.2$~TeV.~\cite{Becirevic:2017jtw}
}
\label{fig:5}
\end{figure}

After applying the constraints described above, we find that this scenario can predict $R_K$ and $R_{K^\ast}$ in good agreement with the findings of LHCb, as illustrated in Fig.~\ref{fig:5}. Interestingly, the $1.5\sigma$ compatibability with $R_{K^{(\ast)}}$ in the central $q^2$-bins imposes the upper bound $m_\Delta<1.2~\mathrm{TeV}$, which can be directly probed at the LHC~\cite{Becirevic:2017jtw}. Furthermore, we found that the branching ratios for the lepton flavor violating (LFV) decays $Z\to \mu\tau$ and $B\to K\mu\tau$ can be as large as $\mathcal{O}(10^{-7})$ and $\mathcal{O}(10^{-9})$, respectively, offering an opportunity for current and future experiments.~\footnote{Note that similiar limits can be derived for the other $b\to s\mu\tau$ exclusive modes by using the relations $\mathcal{B}(B_s\to \mu\tau) \approx 0.9 \times \mathcal{B}(B\to K\mu\tau)$ and $\mathcal{B}(B\to K^\ast\mu\tau) \approx 1.8 \times \mathcal{B}(B\to K\mu\tau)$ derived in Ref.~\cite{Becirevic:2016zri}.}

\section{Conclusions}

In this proceeding we discussed a peculiar form of the $R_2$ model, which postulates the existence of a weak doublet of scalar LQs with hypercharge $Y=7/6$. We showed that this model can explain both $R_{K}^\mathrm{exp}<R_{K}^\mathrm{SM}$ and $R_{K^{\ast}}^\mathrm{exp}<R_{K^\ast}^\mathrm{SM}$ reported by the LHCb collaboration through loop effects of $m_\Delta=\mathcal{O}(1~\mathrm{TeV})$ LQ states. This model has the great advantage of not disturbing the proton stability, and it offers several predictions which can be tested in the near future, such as the bounds $\mathcal{B}(Z\to \mu\tau)\lesssim\mathcal{O}(10^{-7})$ and $\mathcal{B}(B\to K \mu\tau)\lesssim\mathcal{O}(10^{-9})$ for LFV processes. 

\section{Acknowledgments}
This project has received funding from the European Union’s Horizon 2020 research and innovation programme under the Marie Sklodowska-Curie grant agreements No.~690575 and No.~674896.

\end{document}